\documentclass[aps,prl,twocolumn,amsmath,amssymb,showpacs]{revtex4-1}

\usepackage[utf8]{inputenc}
\usepackage{graphicx}
\usepackage{textcomp}
\usepackage{amsmath}
\usepackage{amssymb}
\usepackage{siunitx}
\usepackage{hyperref}

\begin{document}

\title{Coherent control in the ground and optically excited state of an ensemble of erbium dopants}
\author{Pablo Cova Fari\~{n}a}
\email{P.C.F. and B.M. contributed equally to this work.}
\author{Benjamin Merkel}
\email{P.C.F. and B.M. contributed equally to this work.}
\author{Natalia Herrera Valencia}
\email{Present address: Institute of Photonics and Quantum Sciences, Heriot-Watt University, Edinburgh, UK}
\author{Penghong Yu}
\author{Alexander Ulanowski}
\author{Andreas Reiserer}
\email{andreas.reiserer@mpq.mpg.de}

\affiliation{Quantum Networks Group, Max-Planck-Institut f\"ur Quantenoptik, Hans-Kopfermann-Strasse 1, D-85748 Garching, Germany}
\affiliation{Munich Center for Quantum Science and Technology (MCQST), Ludwig-Maximilians-Universit\"at
M\"unchen, Fakult\"at f\"ur Physik, Schellingstr. 4, D-80799 M\"unchen, Germany}

\begin{abstract}
Ensembles of erbium dopants can realize quantum memories and frequency converters that operate in the minimal-loss wavelength band of fiber optical communication. Their operation requires the initialization, coherent control and readout of the electronic spin state. In this work, we use a split-ring microwave resonator to demonstrate such control in both the ground and optically excited state. The presented techniques can also be applied to other combinations of dopant and host, and may facilitate the development of new quantum memory protocols and sensing schemes.
\end{abstract}

\maketitle

In recent years, the spin of optically interfaced dopants and impurities in semiconducting hosts has emerged as a promising platform for quantum technology \cite{awschalom_quantum_2018}. Among all dopants studied to date \cite{thiel_rare-earth-doped_2011}, erbium stands out as it offers an optical transition with exceptional coherence \cite{bottger_optical_2006, merkel_coherent_2020} in the C-band of fiber-optical telecommunication, where loss is minimal. Pioneering experiments have demonstrated the storage of photons \cite{lauritzen_telecommunication-wavelength_2010, probst_anisotropic_2013, dajczgewand_optical_2015}, the integration into waveguides \cite{thiel_rare-earth-doped_2012, miyazono_coupling_2016, askarani_persistent_2020, weiss_erbium_2021} and resonators \cite{probst_anisotropic_2013, chen_coupling_2016, dibos_atomic_2018, chen_parallel_2020, merkel_coherent_2020, casabone_dynamic_2020}, and the potential for second-long hyperfine coherence at high magnetic fields using the isotope $^{167}\text{Er}$ \cite{rancic_coherence_2018}.

In this work, we focus on the electronic spin of the even isotopes of erbium that have no nuclear spin and a natural abundance of $\sim 80\,\%$. Because of the short spin and long optical lifetime, efficient spin initialization has proven difficult in this system and has only been achieved by stimulated emission with an additional laser \cite{lauritzen_state_2008}. Furthermore, coherent control of the electronic spin of erbium ensembles by microwave (MW) pulses has not been demonstrated, neither in the ground nor in the excited state. While the former is typically employed in quantum experiments, the coherence of the latter has only been investigated recently by Raman-Heterodyne spectroscopy \cite{welinski_electron_2019}. Our first demonstration of coherent control on such a microwave transition in the excited state manifold offers interesting perspectives for novel microwave-to-optical transduction and quantum memory protocols \cite{welinski_electron_2019}.

Coherent control of a spin ensemble requires pulses with a bandwidth that exceeds the inhomogeneous broadening. In erbium-doped solids, the spin transition is broadened by around $10\,\si{\mega\hertz}$ in all host materials studied to date \cite{thiel_rare-earth-doped_2011}. Achieving homogeneous MW fields of sufficient strength that can be switched on short timescales has thus been an open experimental challenge.

In this work, we demonstrate that such fields can be generated using a split-ring resonator on a printed circuit board. Our experiments use yttrium-orthosilicate (YSO), a well-studied host for rare-earth dopants \cite{liu_spectroscopic_2005}. In this material, the Kramers' ion erbium substitutes for yttrium in two crystallographic sites \cite{bottger_optical_2006}, each of which has two magnetically inequivalent classes \cite{sun_magnetic_2008}. Our study is performed at a temperature of $0.8\,\si{\kelvin}$, where only the lowest crystal field level of the $I_{15/2}$ manifold of each site and class is occupied. A small magnetic field lifts the two-fold degeneracy of the crystal field levels, which can then be treated as effective two-level systems \cite{bottger_optical_2006}.

As observed earlier, even at low concentrations the lifetime of their spin state is limited by flip-flop interactions \cite{lauritzen_telecommunication-wavelength_2010}, with a strong dependence on the magnetic field orientation \cite{car_optical_2019} that originates from the anisotropic Zeeman Hamiltonian \cite{sun_magnetic_2008}. As the lifetime  improves quadratically with the dopant concentration, we use crystals with a comparably low erbium concentration of $10\,\text{ppm}$, which reduces to $\sim 2\,\text{ppm}$ per site and class for the even isotopes. We use two samples with $0.5\,\si{\milli\meter}$ thickness, but different crystal cut, which exhibit an absorption of around $4\,\%$ when the magnetic classes of the erbium dopants are aligned. Our measurements therefore require averaging of several hundred repetitions per data point. To overcome this difficulty, in future experiments the crystals could be embedded into high-finesse optical resonators \cite{merkel_coherent_2020}, which are directly compatible with the presented setup design.

To initialize and measure the spin state, we drive optical transitions from the ground state to the lowest level of the $I_{13/2}$ manifold, which shows a moderate inhomogeneous broadening of $\sim 300\,\si{\mega\hertz}$ \cite{thiel_rare-earth-doped_2011}. We focus on spins of site~1 with a transition wavelength of $1536.4\,\si{\nano\meter}$, which we drive using laser fields that are switched with two fiber-coupled acousto-optical modulators (Gooch \& Housego). We use commercial laser systems with narrow linewidth (OeWaves Gen3, Koheras Basic X15, or Toptica DLpro) that we can stabilize to a constant frequency difference using home-built beat-detection and locking electronics. 

\begin{figure}[tb]\centering
\includegraphics[width=\columnwidth]{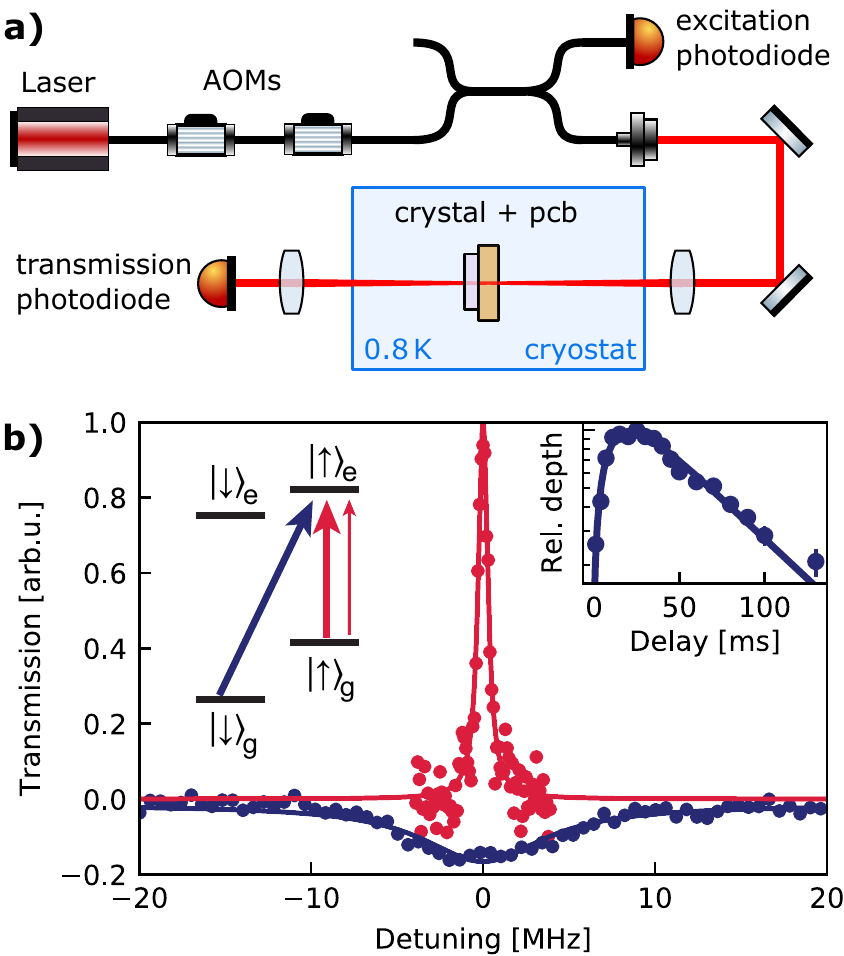}
\caption{\label{fig_HoleBurning}
\textbf{a) Sketch of the optical setup}. The laser is gated and frequency-tuned by fiber-coupled acousto-optical modulators (AOMs). It is focused onto the crystal, which is kept in a closed-cycle cryostat in direct proximity to a MW resonator on a printed circuit board (pcb). The transmission and excitation power are measured by photodiodes. 
\textbf{b) State initialization and readout}. As indicated by arrows in the level scheme (top left), we measure the transmission of the crystal (faint red arrow) after irradiating a burn laser pulse on the spin preserving (red arrow, data and fit) or spin-flip (blue arrow, data and fit) transition. Inset: After burning on the spin-flip transition, the antihole depth shows  biexponential dynamics (blue data and fit curve). It first increases within the optical lifetime and then decays within the spin lifetime.
}
\end{figure}

The laser beams are delivered to the sample either by free space optics or by polarization-maintaining and single-mode optical fibers, as shown in Fig.~\ref{fig_HoleBurning}(a). In the latter case, they are collimated in the cryostat using anti-reflection coated graded-index-lenses (Thorlabs GRIN2915) that are clamped to the same V-grove that also holds a fiber in its ferrule (SMPF0215), achieving an overall transmission around $10\,\%$. To ensure sufficient signal-to-noise ratio, the transmission is measured by an avalanche photodiode (Thorlabs PDB570C). The influence of fluctuations of the laser power is eliminated by dividing the transmission by the independently measured input power, or by using a heterodyne detection technique. The signals are digitized by a real-time experimental control system (NI CompactRIO) that also switches the lasers.

In the following, we present two schemes that enable the initialization and readout of rare-earth spin ensembles in the ground and excited state manifold, starting with the former. Previous experiments at zero magnetic field found that it is not possible to achieve efficient spin pumping in Er:YSO \cite{lauritzen_state_2008}. We overcome this challenge by applying magnetic fields of about $0.02 \,\si{\tesla}$ along the D2-axis of YSO, which splits the ground state spins by $3.12\,\si{\giga\hertz}$ and the excited states by $2\,\si{\giga\hertz}$. The magnetic field is generated by two neodymium disk magnets (Maqna) with 7 cm diameter, located outside of the cryostat on micrometer stages, which allows for precise alignment of the field amplitude and direction. Thermal drifts lead to a change of the magnetic field on the order of $10^{-3}$ over several weeks, which is smaller than the inhomogeneous broadening of our spin ensembles and thus does not affect the measurements.

With the magnetic field, the optical spin-flip transitions are detuned from the spin-preserving ones by approximately $2.5\,\si{\giga\hertz}$, which exceeds the inhomogeneous broadening of about $0.5\,\si{\giga\hertz}$. When selectively driving the spin-flip transition for $100\,\si{\milli\second}$, we therefore observe an antihole in the transmission at the center of the spin preserving line, as shown in fig.~\ref{fig_HoleBurning}b (blue data and fit). The height of the antihole is directly proportional to the net polarization of the ground state spins, which allows for spin readout in coherent control experiments.

Measuring the decay of the antihole over time (inset), allows us to determine the lifetime of the optical and spin transition. We find a value of $53(3)\,\si{\milli\second}$ for the latter, limited by flip-flop interactions (as observed and characterized recently under similar conditions \cite{car_optical_2019}). The ground-state lifetime is thus considerably longer than that of the excited state, $11\,\si{\milli\second}$ \cite{bottger_spectroscopy_2006}. Thus, even without stimulated emission using another laser \cite{lauritzen_state_2008} or optical resonator \cite{ miyazono_coupling_2016, dibos_atomic_2018, casabone_cavity-enhanced_2018, casabone_dynamic_2020, merkel_coherent_2020}, by applying a sufficient magnetic field we can achieve efficient spin initialization. 

\begin{figure*}[tb]
\includegraphics[width=\linewidth]{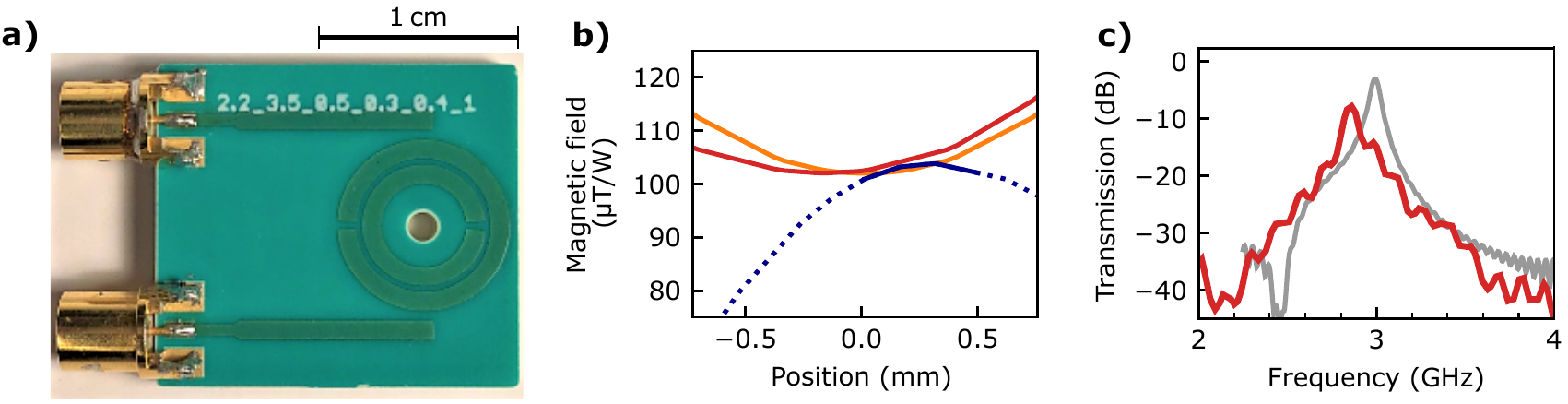}
\caption{\label{fig_Resonator}
\textbf{Details of the microwave resonator.} (a) Photograph of the MW resonator. Two striplines capacitively couple to the resonator that consists of two concentric split-rings. (b) Simulated magnetic field homogeneity along the laser propagation axis (solid blue inside crystal, dotted blue in vacuum), as well as perpendicular to it (red and orange). In the probed sample volume of $\sim 0.2\times0.2\times0.5\,\si{\milli\meter}^3$, the MW field amplitude changes by about $2\,\%$. (c) MW transmission from the input to the output port of the resonator. The measurement (red) shows good agreement with the simulation (grey). Excess loss of $\sim 5\,\text{dB}$ and the slight oscillation is attributed to the loss and impedance mismatch of the cables and components.
}
\end{figure*}

Using a rate equation model based on the experimental timescales of the hole and antihole decays and the branching ratio of the excited state decay to the two ground states \cite{petit_demonstration_2020}, we calculate a spin pumping  efficiency around $0.9(1)$. Within errors, this value agrees with our experimental data when comparing the area of the antihole (blue data and Lorentzian fit) with that of a hole burnt on the spin-preserving transition (red arrow, data and Lorentzian fit). While the width of the latter is limited by spectral diffusion \cite{bottger_optical_2006, bottger_effects_2009} and power broadening, the former is set by the inhomogeneous broadening of the spin transition. We find a full-width-half-maximum linewidth of $\sim9\,\si{\mega\hertz}$.

After describing spin initialization in the ground state, we now turn to the optically excited state. Here, one can use a single laser resonant with the spin-preserving transition both for state readout and preparation. Compared to optical pumping, this state initialization technique has two advantages: First, it does not require a large splitting of the spin states, which allows to freely choose the qubit transition frequency. Second, the initialization can be very quick, typically $\sim 1\,\si{\micro\second}$, as it is only limited by the optical Rabi frequency. Therefore, this scheme can be applied even at elevated temperature, or in highly-doped crystals in which the flip-flop dominated ground-state spin relaxation is much faster than spontaneous decay from the excited state \cite{car_optical_2019}, such that optical pumping is not possible.

In our experiments, the width of the spectral hole is narrow ($\lesssim 1\,\si{\mega\hertz}$) and thus only a small fraction of the optical inhomogeneous linewidth is probed. Still, as there is no correlation between the broadening in the ground and excited states, the spin transition linewidth of both manifolds is comparable. Thus, coherent control in the excited and ground state pose the same challenging requirements on the amplitude and bandwidth of MW control fields. In contrast to measurements with single emitters \cite{de_lange_universal_2010, kornher_sensing_2020, chen_parallel_2020} or small ensembles \cite{zhou_quantum_2020} that can use close-by striplines, experiments with extended ensembles thus require tailored resonators. Previous setups with  erbium \cite{fernandez-gonzalvo_coherent_2015, chen_coupling_2016} and NV center ensembles \cite{angerer_collective_2016} used high-Q resonators. This has the advantage of requiring little drive power, but inevitably comes at the price of a small bandwidth -- too small to achieve coherent control of the ensemble on the sub-$\si{\micro\second}$ timescale of the spin bath evolution in YSO crystals with Kramers dopants \cite{kornher_sensing_2020, merkel_dynamical_2020}.

To overcome this challenge, we use a resonator with moderate $Q \simeq 50$, as shown in Fig.~\ref{fig_Resonator}(a). It follows the design used in \cite{bayat_efficient_2014}, with a slight modification of the geometry to account for the different resonance frequency and substrate (Rogers~4350B). With sufficient input power, it achieves a strong ($\sim 100 \si{\micro\tesla}/\si{\watt}$) and homogeneous (variation $\lesssim 2\,\%$) magnetic field over an extended sample volume of approximately $(0.5\,\si{\milli\metre})^3$, as shown in (b). Albeit the mode volume of the resonator, $\simeq 10^{-5}\,\lambda^3$, is small compared to the vacuum wavelength $\lambda$, the system does not reach the strong-coupling regime \cite{probst_anisotropic_2013} because of the moderate $Q$.

A transmission measurement shows a resonance at $3.12\,\si{\giga\hertz}$ with a linewidth of $60(1)\,\si{\mega\hertz}$  (c, red curve), which is close to the designed frequency and width according to our simulations (grey). The bandwidth thus exceeds the inhomogeneous broadening of the spin transition, which allows for fast pulses that drive the entire ensemble.

To this end, we generate MW signals by applying pulses from an arbitrary waveform generator (Zurich Instruments HDAWG) to the pulse and IQ-modulation input of a continuous-wave source (Rohde and Schwarz SGS100A). The MW signal is amplified (Mini Circuits ZHL-100W-352+) to a peak power of $0.1\,\si{\kilo\watt}$. The MW resonator is thermally anchored at the ground plane side to an oxygen-free copper block. In pulsed measurements at full power, using a close-by resistive thermometer we observe a heating of $0.05\,\si{\kelvin}/\si{\milli\watt}$ (cw), which limits the number of pulses that can be applied to the sample. 

\begin{figure}[tb]
\centering
\includegraphics{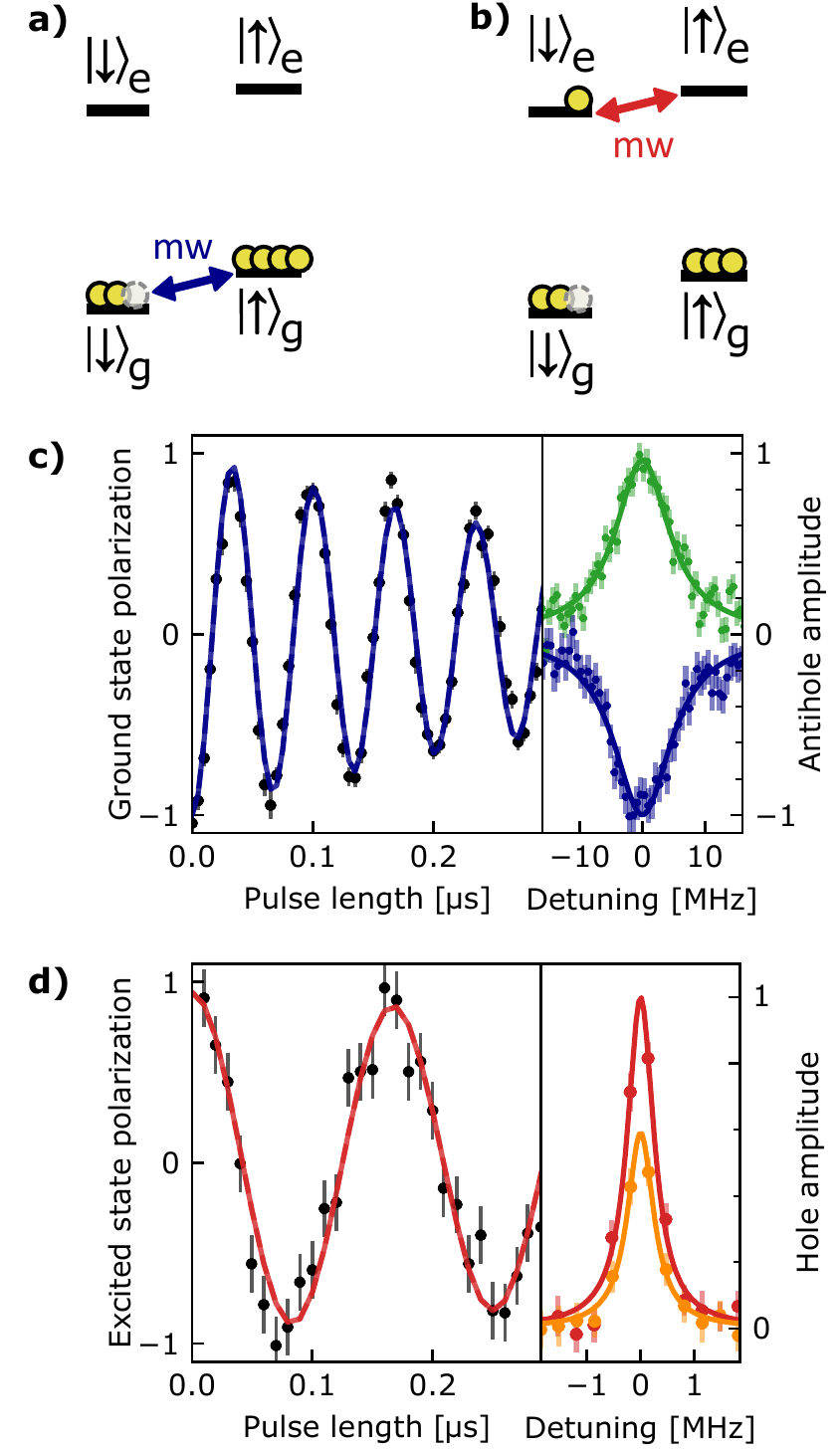}
\caption{\label{fig_MW_control}
\textbf{Coherent control of erbium ensembles.} We apply $0.1\,\si{\kilo\watt}$ MW pulses to the $3.12\,\si{\giga\hertz}$ resonance of a double split-ring resonator, which drives Rabi oscillations in the ground level (a,c) or in the excited level (b,d). 
In the ground-state experiments, the high Rabi frequency of $2\pi\cdot14.9(5)\,\si{\mega\hertz}$ enables complete inversion over a large fraction of the inhomogeneous distribution, as can be seen by comparing the amplitude of the antihole measured before (c, right panel, blue) and after (green) applying a $33\,\si{\nano\second}$ long $\pi$-pulse. In experiments on the excited state, we measure a lower Rabi frequency of $2\pi\cdot6.2(2)\,\si{\mega\hertz}$, because of a reduced transition strength in this magnetic field configuration. The maximum contrast in the hole depth before (d, right panel, red) and after (yellow) a $\pi$-pulse approaches the ideal value of $0.5$.
}
\end{figure}

We now use these MW pulses to drive an ensemble of erbium spins after the initialization procedure described above. When varying the pulse duration, we observe Rabi oscillations in both the ground and excited state manifolds, as depicted in the left panel of Fig.~\ref{fig_MW_control}(c) and (d), respectively. After $500\,\si{\nano\second}$ of continuous MW irradiation, we observe a reduction of the Rabi frequency by $\sim 10\,\%$ (not shown), which we attribute to an increased cable attenuation caused by heating.

The interaction of erbium dopants with magnetic fields can be modeled as effective spin-1/2 electron with an anisotropic g-tensor. Both, the Zeeman splitting and the spin transition strength depend strongly on the orientation of the static and MW magnetic fields and are characterized by effective g-factors \cite{maryasov_spin_2012, probst_anisotropic_2013}. 
For experiments on ground-state spins, we oriented the static magnetic field along the D2-axis of YSO and the MW field along the b-axis. This configuration has been chosen as a compromise between a large ground-state g-factor of along the magnetic field, $g_g=10.5$ (to ensure sufficient lifetime), and along the MW direction, $g_\text{mw}=1.6$ (to enable fast pulses). 
We obtain a Rabi frequency of $2\pi\cdot14.9(5)\,\si{\mega\hertz}$ (blue fit curve), which allows for an almost complete inversion of the population after a $\pi$-pulse of $33\,\si{\nano\second}$ length (right panel). At the line center we measure a $\pi$-pulse fidelity of $98\,\%$, and its average over the full inhomogeneous linewidth is $0.78(6)$ following the analysis of \cite{agnello_instantaneous_2001}. 

In experiments on optically excited spins, we aligned the static magnetic field along the b-axis and the MW field along D2. Again, this configuration is a compromise between a large-enough g-factor parallel to the MW, $g_\text{mw}=0.95$ (for fast pulses), and parallel to the external field direction, $g_e=10$ (to allow tuning the transition into resonance with the MW using the magnetic fields achievable in our setup). We measure a Rabi frequency of $2\pi\cdot 6.2(2)\,\si{\mega\hertz}$ (red fit curve). The reduction is explained by the smaller effective g-factor for MW fields in this orientation and slightly higher MW insertion loss obtained in this resonator assembly compared to the previous one.

In ground-state experiments, the signal changes from an antihole after initialization (fig.~\ref{fig_MW_control}c, right panel, blue) to a hole of the same depth after a $\pi$-pulse (green). In contrast, even in case of perfect pulses the maximum signal difference in the optical readout within the excited-state manifold is smaller (fig.~\ref{fig_MW_control}d, right panel, red and yellow). The reason is that the spectral hole is produced by two effects: the reduction of the absorption from the ground state and the increase of stimulated emission from the excited state. After a $\pi$-pulse in the excited state, the probe laser can no longer stimulate emission, but the absorption stays reduced. Therefore, the maximum contrast after perfect inversion would be $0.5$. 

The above considerations allow us to calibrate the spin polarization and extract the MW pulse quality. In both the ground and excited state, we obtain $\pi$-pulse fidelities of $98\,\%$, in good agreement with our expectation from the simulated field inhomogeneity of the MW resonator. If required, further improvement of the control fidelity can be obtained by composite or shaped pulses \cite{vandersypen_nmr_2005}.

In addition to the imperfect field homogeneity, also the coherence of the spin ensembles may limit the achievable control fidelity. In Ramsey measurements, we find coherence times of $0.056(9)\,\si{\micro\second}$ and $0.07(1)\,\si{\micro\second}$ in the ground and excited states, respectively. Within errors, the obtained values are identical and limited by the inhomogeneous broadening of the spin transition. In contrast, the coherence of a spin-echo measurement differs between the manifolds. In the excited state, it is limited by the interaction with the nuclear spin bath, whereas in the ground state it is limited by the interaction within the resonant erbium spin ensemble. A detailed study of these effects is presented in \cite{merkel_dynamical_2020}.

To summarize, we have demonstrated the intialization, coherent control and readout of ensembles of erbium dopants in YSO. The achieved pulse fidelities allow for detailed studies of the spin coherence and its improvement by dynamical decoupling \cite{merkel_dynamical_2020}. Furthermore, our measurements pave the way to use Er:YSO as a novel platform for quantum sensing. Compared to established systems such as the NV-center in diamond \cite{zhou_quantum_2020} and most other impurities \cite{awschalom_quantum_2018}, a sevenfold larger sensitivity to magnetic fields is obtained by the high effective g-factor of erbium dopants. Finally, our demonstration of coherent control in the excited state of rare-earth dopants may allow for the implementation of recently-proposed protocols for quantum transduction and memory \cite{welinski_electron_2019}.

\section{acknowledgements}

We acknowledge the contribution of Kutlu Kutluer during an early stage of the project. This project received funding from the European Research Council (ERC) under the European Union's Horizon 2020 research and innovation programme (grant agreement No 757772), from the Deutsche Forschungsgemeinschaft (DFG, German Research Foundation) under Germany's Excellence Strategy - EXC-2111 - 390814868, and from the Daimler-and-Benz-Foundation.

\bibliography{bibliography.bib}

\end{document}